# APPSECURE.NRW SOFTWARE SECURITY STUDY[1]

## A survey among developers, product owners, and managers

Stefan Dziwok, Thorsten Koch, Sven Merschjohann, Boris Budweg, Sebastian Leuer
*Fraunhofer IEM, Germany*
*{firstname.name}@iem.fraunhofer.de*

**Abstract**: In recent years, the World Economic Forum has identified software security as the most significant technological risk to the world's population, as software-intensive systems process critical data and provide critical services. This raises the question of the extent to which German companies are addressing software security in developing and operating their software products. This paper reports on the results of an extensive study among developers, product owners, and managers to answer this question. Our results show that ensuring security is a multi-faceted challenge for companies, involving low awareness, inaccurate self-assessment, and a lack of competence on the topic of secure software development among all stakeholders. The current situation in software development is therefore detrimental to the security of software products in the medium and long term.

---

[1] This document has been translated from German. The original version is available at https://www.appsecure.nrw/wp-content/uploads/2021/03/AppSecure_nrw_Studie.pdf

# Table of Contents

# 1. Introduction

*"Something is happening. It is gaining importance but is not accepted by everyone with a kiss because security still has a braking effect. [...] I don't think anyone would say that security doesn't matter, certainly not! But it is exhausting to deal with the topic. "*

The security requirements for software-intensive systems are constantly rising as software-intensive systems increasingly process data requiring protection and provide critical services. Therefore, it is not surprising that the World Economic Forum, in its Global Risk Report 2020, classifies security as the most significant technological risk for the world's population. This raises the question to which extent German companies address security in developing and operating their software products. We conducted an extensive study in our research project AppSecure.nrw last year to answer this question. We surveyed the software developers and their managers and product owners to obtain a holistic picture.

The result of our study is that ensuring security in their products is a multi-layered challenge for companies, and there is a need for action. Primarily, developers, managers, and product owners lack awareness regarding security, security expertise, methods for secure software development, and suitable tools for this purpose. Furthermore, we have recognized that the product owners place too few or no security requirements on the product to be developed. Additionally, the managers only rarely implement measures to increase security competence. Thus, many companies risk that their products are not protected against malicious attacks. Fortunately, many developers, managers, and product owners know their current challenges and are willing to improve the current situation. We support this improvement in the further course of AppSecure.nrw by defining a maturity model for agile teams, creating additional training for developers, managers, and product owners, and further developing existing free tools.

In the following, we present our study results in detail. After an explanation of our methodology and the professional background of the study participants, we present the central results of our study based on five topics: First, we focus on the development process, which we have divided into the disciplines of requirements engineering, design, implementation & testing, and operation. We then report on our findings on the use of tools. After that, we address the security competence of all participants and the current measures for competence development. In the last topic, we analyze security awareness. Finally, we conclude our study and outline its effects and recommendations to improve the secure software development.

We wish you a stimulating read. If you have any questions about the study or our recommendations, please do not hesitate to contact us.

## 2. Methodology

We developed an online survey and two interview guides based on 21 research questions using qualitative and quantitative research methods. Afterward, we continuously improved the comprehensibility and clarity of the online survey together with the AppSecure.nrw project partners AXA Konzern AG, Connext Communication GmbH, and adesso mobile solutions GmbH. Fraunhofer IEM is responsible for developing, implementing, and evaluating the study.

The 40-question online survey was conducted anonymously and promoted by Fraunhofer IEM and its business partners, the AppSecure.nrw project partners, the technology networks it's OWL and innozent OWL, Heise Medien, and Bitkom. A total of 350 people took part in the survey. Unfortunately, we had to exclude responses from Austria and Switzerland, as they participated in a too-small number for a meaningful evaluation. Furthermore, we filtered out all participants who did not complete the survey in full or whose processing time did not allow them to read the questions. Finally, we evaluated a total of 256 respondents from Germany. This selection required an average processing time of 25 minutes to answer all questions. Concerning our presentation of results, we rounded all percentages to whole numbers. Therefore, all deviations of the sum from 100 percent are due to rounding.

Four of the 17 people interviewed are part of the AppSecure.nrw project team. The remaining 13 people have no connection to the project. The interview guides for managers and product owners consisted of 17 questions. If an interviewee held both roles, we asked the questions from both guides. Follow-up questions and digressions were also deliberately allowed. On average, an interview lasted about 45 minutes. The interviews showed a significant overlap between the two roles manager and product owner. Therefore, we often analyzed both roles together. However, if there were differences, we always marked them. All quotes that we present in this document originate from these interviews.

# 3. The Professional Background of the Participants

In this chapter, we present the professional background of the interviewed developers, managers, and product owners.

## 3.1. Developers

Two hundred fifty-six developers from Germany completed the online survey. 61% state that they have more than ten years of professional experience in software development (cf. Figure 1). 18% of the developers have between six and ten years of professional experience, and 15% have between two and five years of professional experience. 5% of the developers are newcomers to the profession and have less than two years of professional experience. Overall, we mainly reached very experienced developers with our online survey.

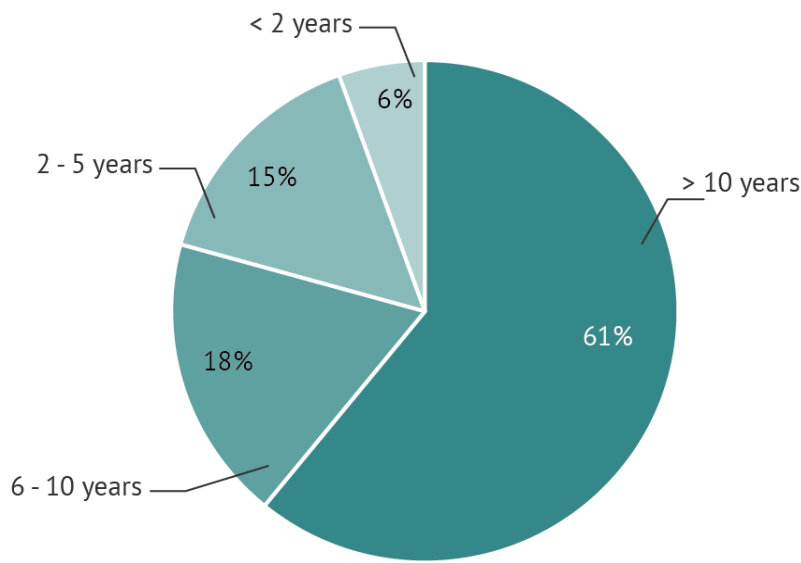

*Figure 1: Overview of the professional background of the developers interviewed*

Broken down by company size (cf. Figure 2), 14% of the developers come from companies with a maximum of 50 employees. 26% come from companies with at least 51 and a maximum of 250 employees. Overall, 40% of the developers work at small and medium-sized enterprises (SMEs). On the other hand, 61% of the developers work in large companies (more than 250 employees).

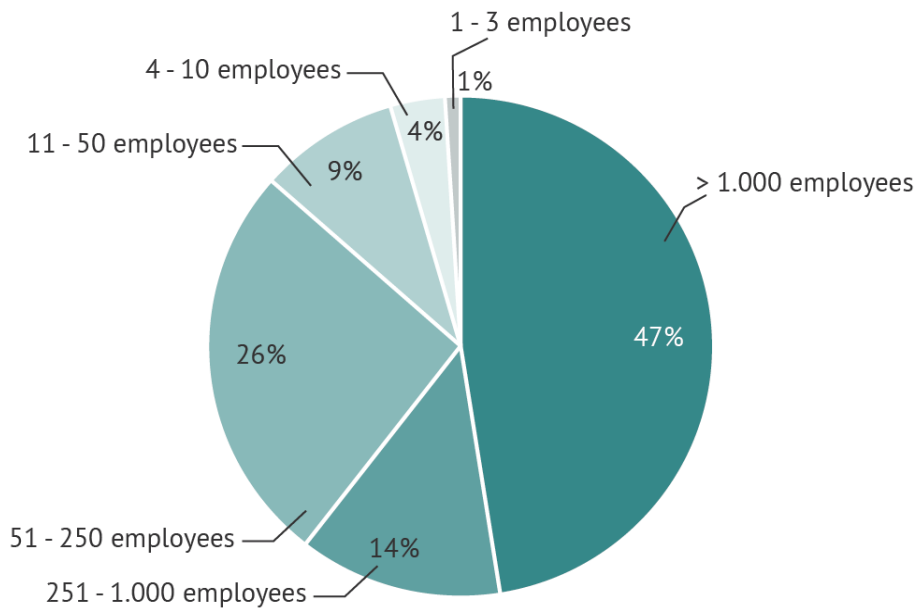

*Figure 2: Company size of the developers surveyed*

The companies' business models are very different (cf. Figure 3). In most cases, the software (55%) is used within the company. The software is also licensed to customers (36%) or developed directly for customers (28%). 15% of the developers are also sent to other companies to work on projects there.

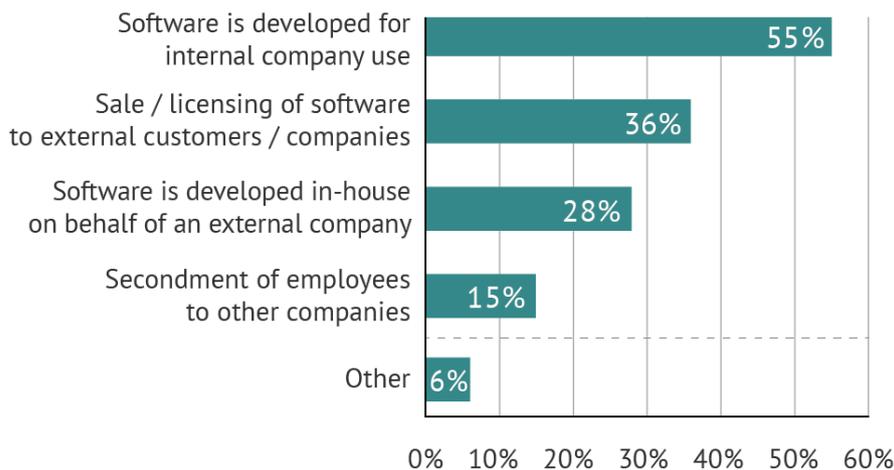

*Figure 3: Business models of the companies*

The developers are involved in various applications (cf. Figure 4). Web applications (66%) and backend applications (59%) are developed particularly frequently, followed by desktop applications (37%) and mobile applications (25%). On the other hand, applications for the embedded environment are developed relatively rarely (14%).

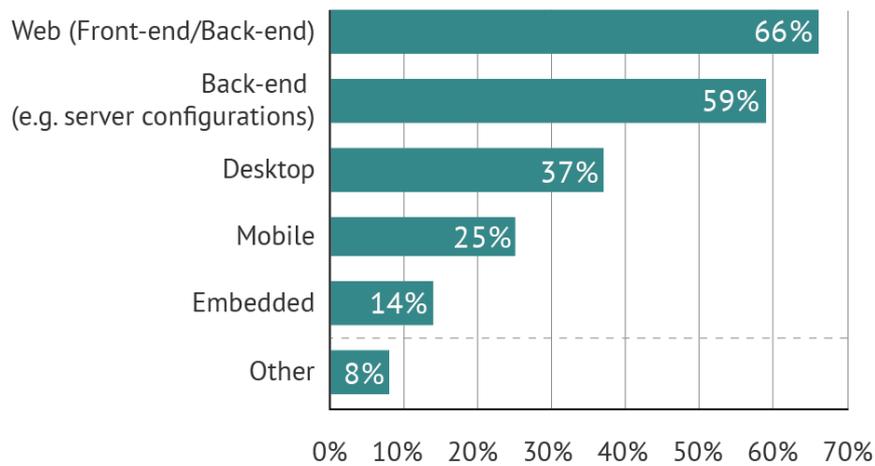

*Figure 4: Application types of the developers*

The size of the teams among the developers is usually 15 or fewer people (cf. Figure 5). The developers typically work in teams of 6-15 people (56%) or 1-5 people (32%). Larger teams with more than 16 people are relatively rare (13%).

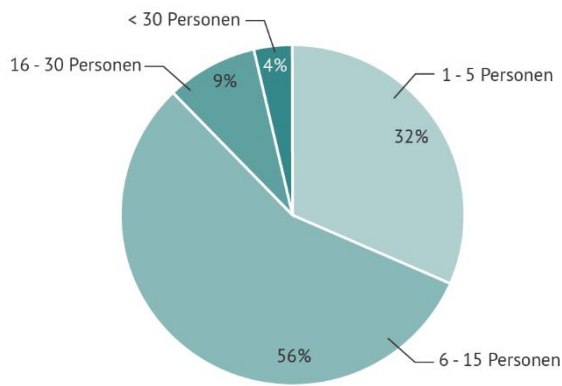

*Figure 5: Team size of the developers*

Developers typically work in several disciplines: 64% are active in requirements engineering, 81% in design, 86% in implementation or software testing, and 52% in operating the software (cf. Figure 6).

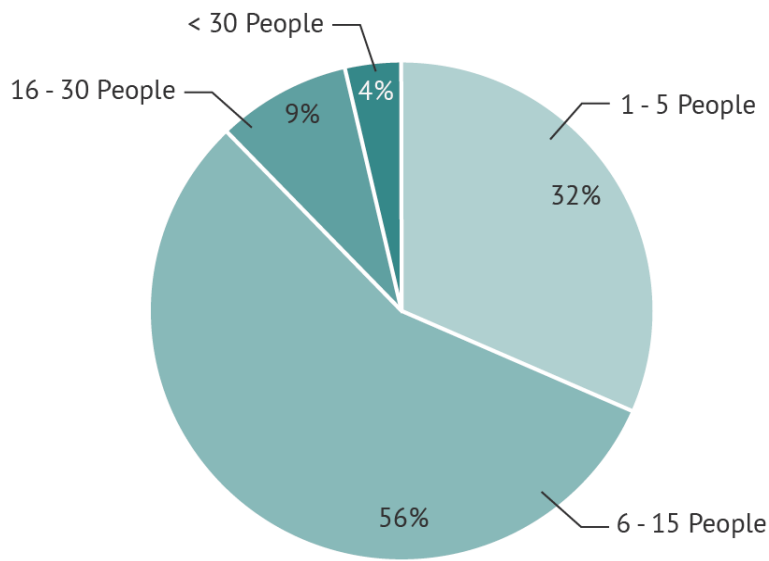

Question: "What is the size of the team you (typically) work on?"
N=256.

*Figure 6: Proportion of the developers per discipline*

The developers in our study use a wide range of development environments and programming languages (cf. Figure 7). IntelliJ and Eclipse are used particularly frequently. The development environments Visual Studio and Visual Studio Code are also widely used. In terms of programming languages, Java, JavaScript/TypeScript, and C# are in the lead (cf. Figure 8). On average, developers use two to three programming languages and two different development environments.

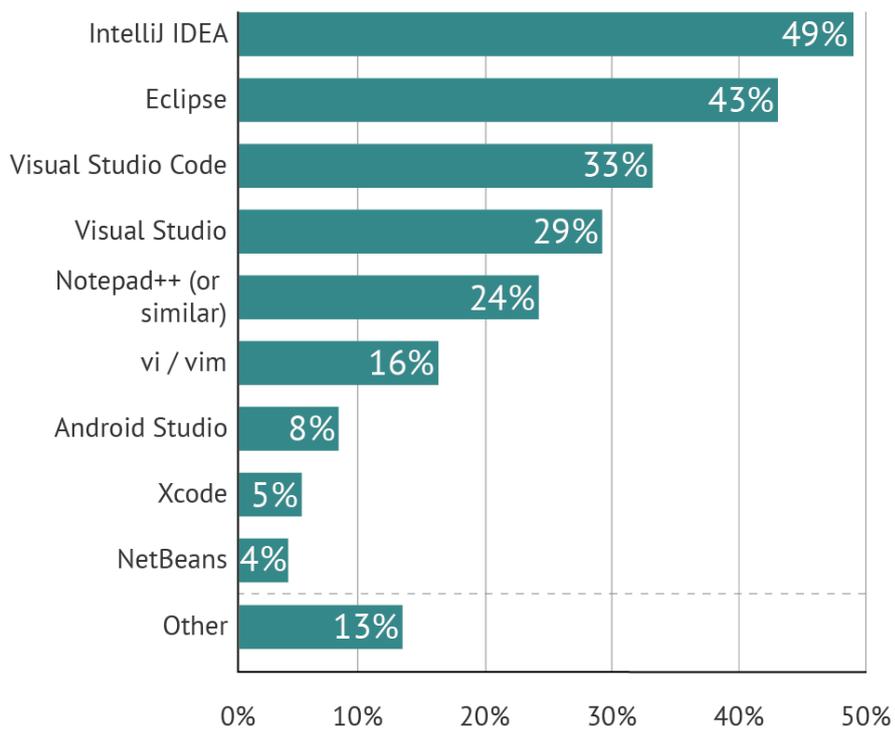

*Figure 7: Overview of the development environments used*

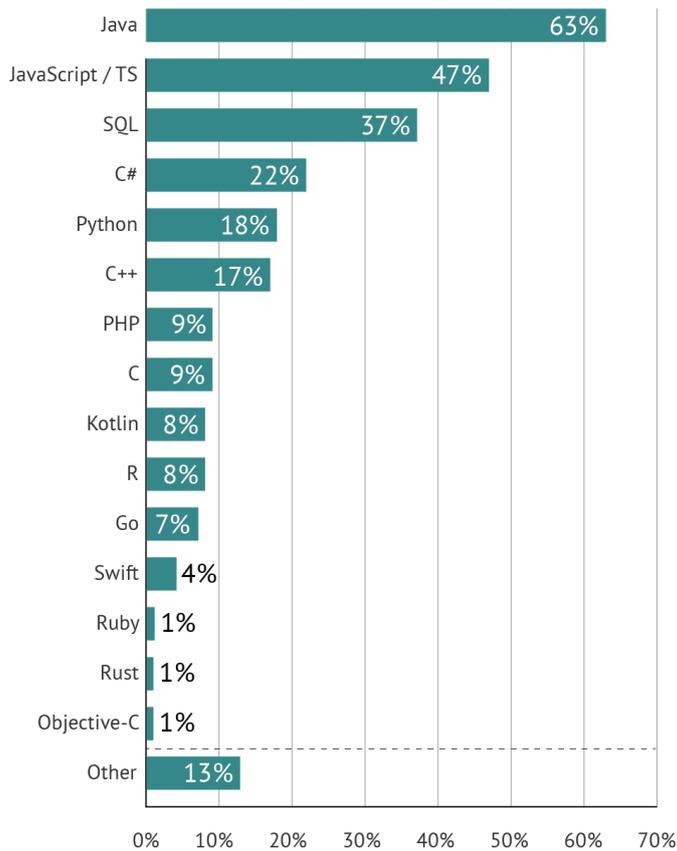

Question: "Which languages do you primarily use in your team?"
Multiple answers possible. N=221

*Figure 8: Overview of the programming languages used*

## 3.2. Managers and Product Owners

We interviewed 17 people using an interview guide. Seven of the interviewees work exclusively as product owners, and six work exclusively as managers. The remaining four individuals hold both roles. All interviewed people have several or many years of professional experience and always have direct contact with software development.

As the developers in the online survey, the managers and product owners interviewed were from both small (2 people) and medium-sized companies (10 people), as well as companies with more than 250 employees (5 people).

The companies for which the surveyed managers and product owners work develop software for various industries (e.g., automotive, healthcare, and insurance). Additionally, almost half of the managers and product owners state that they develop software for more than one industry.

The business models also differ among the companies. Primarily, the software is developed for internal use, direct customer order, or licensed sale. Two interviewees also work for companies that send their employees to other companies.

# 4. Development Process and Operation

> "It's not as if we don't think about security at all. But of course, there is always an expense that the employee then carries with him. So, you always have to look at it in comparison."

In addition to suitable tools and competent developers (cf. Chapters 5 and 6), the development process is crucial for a secure product. Therefore, the idea of the so-called security-by-design approach is the following: Security is considered from the outset and throughout the entire development process. Figure 9 shows a suitable development process[2] in which all disciplines have been supplemented with exemplary security-enhancing methods. The methods are intended to ensure that serious weaknesses are found early in the development process. This significantly reduces the effort required to correct errors. But how present is the topic of security among developers in the four disciplines of requirements engineering, design, implementation & testing, and operation? And how well do developers, product owners, and managers see themselves positioned concerning security in their current processes? We used our study to put light on these three questions. We asked all participants in the study for their general assessment and asked the developers to assess the four disciplines.

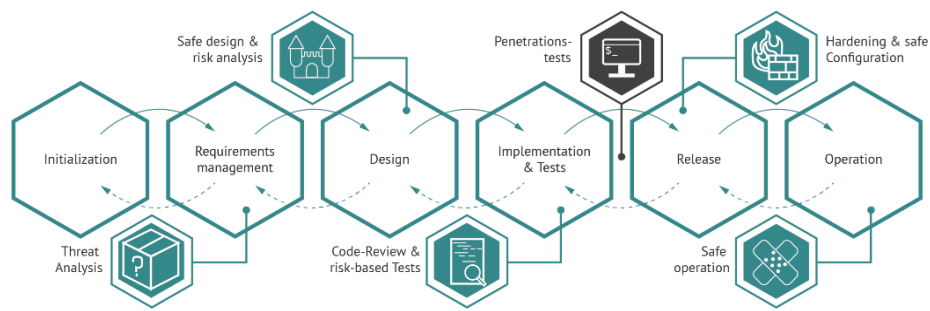

*Figure 9: Security-by-Design development process*

## 4.1. Is Security Systematically Considered in the Development Process?

We asked the developers whether and how they address security in their development process (cf. Figure 10). 59% of the developers state that they do not have clear guidelines and processes for secure software development. Furthermore, 41% of the developers state that their security requirements (e.g., processing sensitive data) are not clearly defined or known.

In addition, 56% of the developers state that their collection of tools is unsuitable. This can be since no suitable tools are available on the market and because better tools cannot or may not be used, or because better tools have not yet been sought.

Moreover, 62% of the developers state that no fixed person in their team is responsible for security. This can either be since security is not explicitly considered or since everyone in the team is supposed to consider security. Therefore, no one is explicitly responsible. Our general recommendation is to assign this responsibility to a specific person. A clear understanding of the task is important to ensure that everyone in the team also pays attention.

---

[2] For the security-by-design development process, it is not relevant whether the product is developed according to waterfall, V-model or agile. In agile development, the phases from requirements engineering to commissioning take place within one or more sprints.

Based on the developers' assessment, security is not systematically considered by many teams, which results in a high risk that they will not be able to guarantee the sustained high quality of their software products.

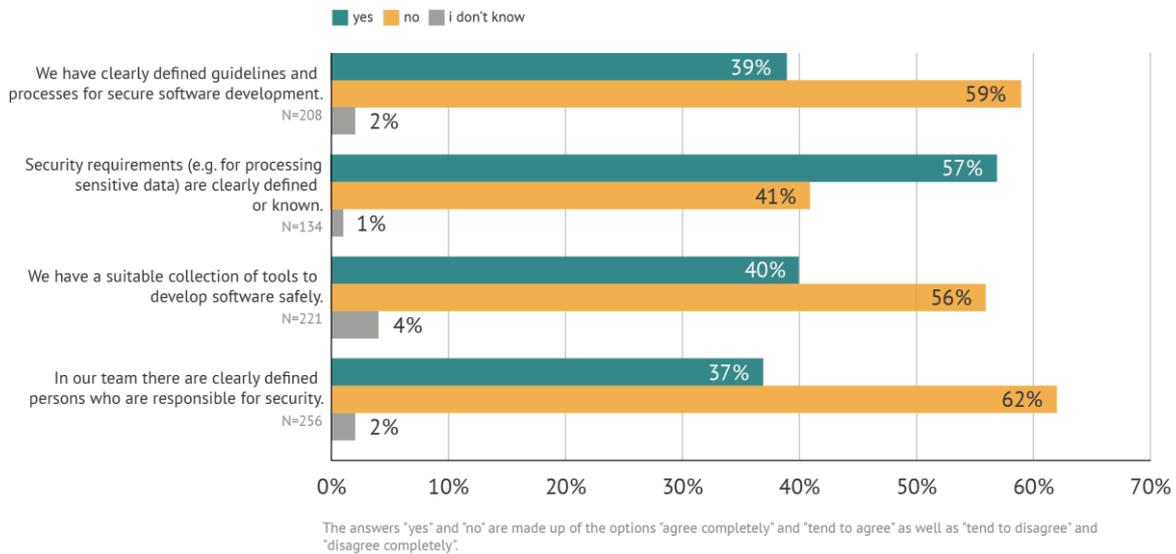

*Figure 10: Consideration of security in the development process*

The interviews with managers and product owners revealed that security has not a high priority in the software development process but rather a low to medium priority. The majority refers to the pen test[3], which only occurs after development. Systematic consideration of the security during the development, for example, using checklists, using a firm assignment of responsibility, or by explicit security guidelines, is only very rarely found. In addition, most product owners state that security plays only a minor or no role at all in the agile meetings (planning, retro, review). Thus, security is typically handled unsystematically in the software development process (i.e., without the establishment of concrete measures), which results in an increased risk for the quality of the resulting products.

"Security usually starts at the end of the whole process. Whereas I always have the feeling that it should begin much earlier. "

## 4.2. Is Security Systematically Considered in the Individual Disciplines?

67% of the developers involved in the discipline of *requirements engineering* state that they pay attention to security (cf. Figure 11). However, only 24% have templates or standards for documenting security requirements. Thus, the requirements seem to be considered unsystematically in many projects. A systematic approach would improve the product because it reduces the risk of formulated incorrectly or forgotten requirements.

Only 35% of the developers have a security expert in their team who checks the security requirements. A designated and trained person would help to ensure that the security requirements are complete and meaningful, thus, laying the foundation for a secure product at an early stage.

---

[3] A penetration test is typically carried out before the release of software - often by a third party. The finished product is executed and checked for vulnerabilities (e.g. incorrect configurations, bad passwords, security against standard attacks).

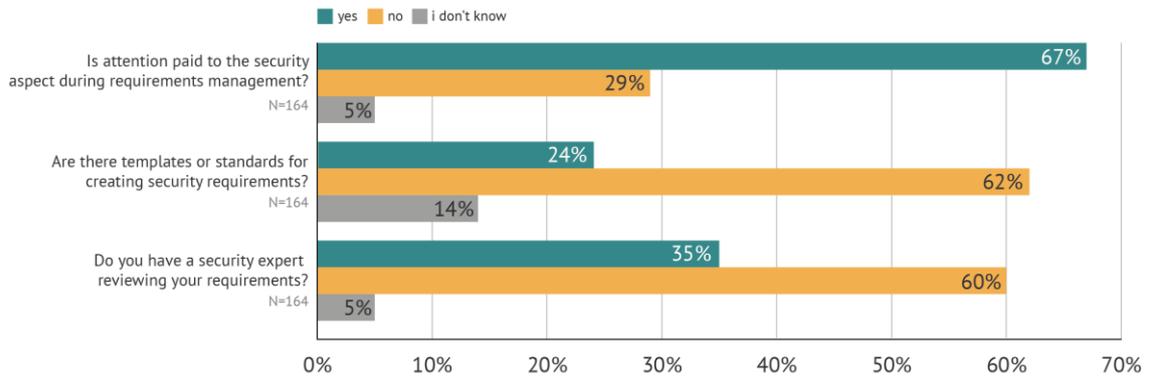

*Figure 11: Consideration of security in the requirements discipline*

The picture for *design* is similar to the requirements engineering (cf. Figure 12). 77% of the developers state that they pay attention to security during the software system design, but only 24% have templates or standards. Again, only 32% of the developers have an expert to check the security of the design.

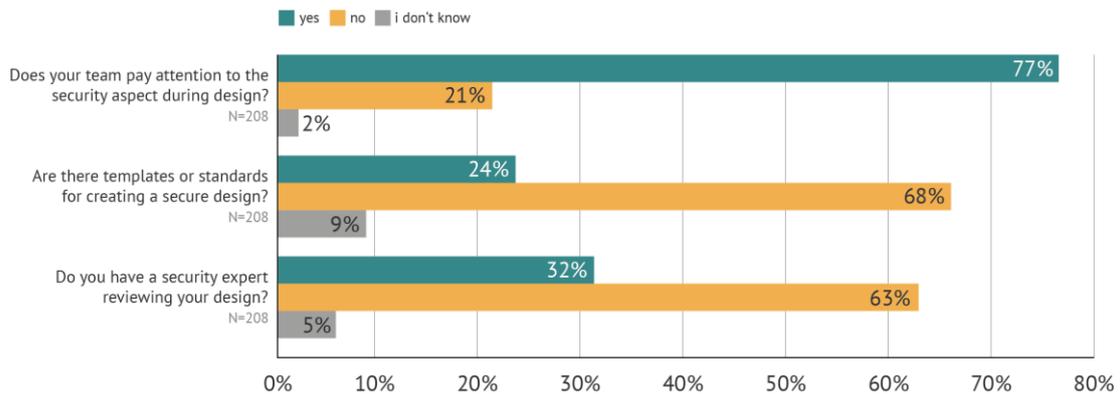

*Figure 12: Consideration of security in the design discipline*

In the *implementation and testing* discipline, 75% of the developers state that they pay attention to security (cf. Figure 13). In this context, 27% of the developers use appropriate templates and standards to assist them. To ensure that all security properties of their developed product are adhered to, 25% of the developers have a corresponding process. However, only 16% of the developers carry out a final security review before a release.

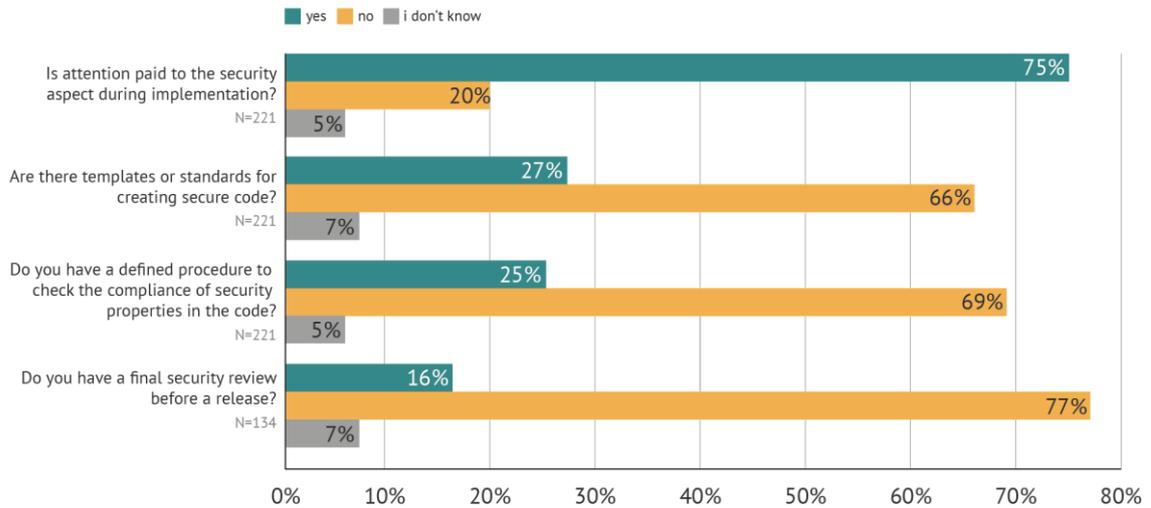

*Figure 13: Consideration of security in the discipline of implementation and testing*

In response to our question about when the software is analyzed concerning security, we allowed several possible answers since repeated analysis is highly recommended (cf. Figure 14). Almost half (48%) of the developers state that this already happens during programming. Another popular time to analyze security is before each release (37%). Other points in time were before check-in (10%), after a check-in (20%), during the sprint (20%), as well as isolated other points in time (7%), such as after the release or when pen tests are carried out. Overall, we determine that most developers analyze their software at least at one point concerning security. However, 20% of the developers also state that they never analyze the security of their software.

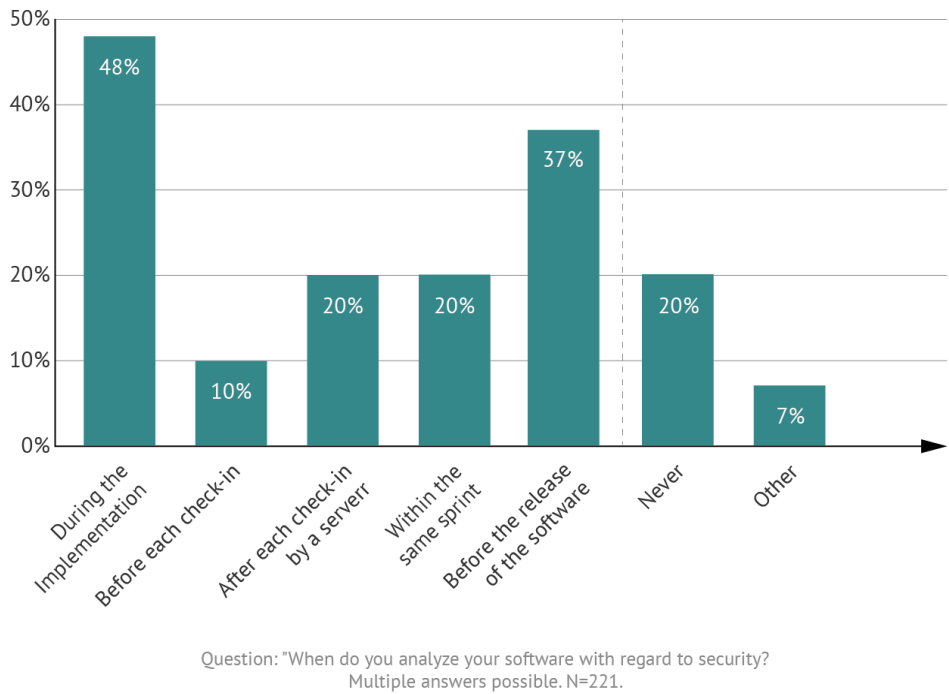

*Figure 14: When is the software analyzed concerning security?*

22% of the developers perform automated security checks while the software is *running* (cf. Figure 15). 28% carry out automatic security checks after a release. If security problems (e.g., concerning libraries used) are found in products in operation, 40% of the developers state that they are aware of this. In addition, 40% of the developers also say that they have clear guidelines for dealing with (potential) security vulnerabilities, attacks, data leaks in operation.

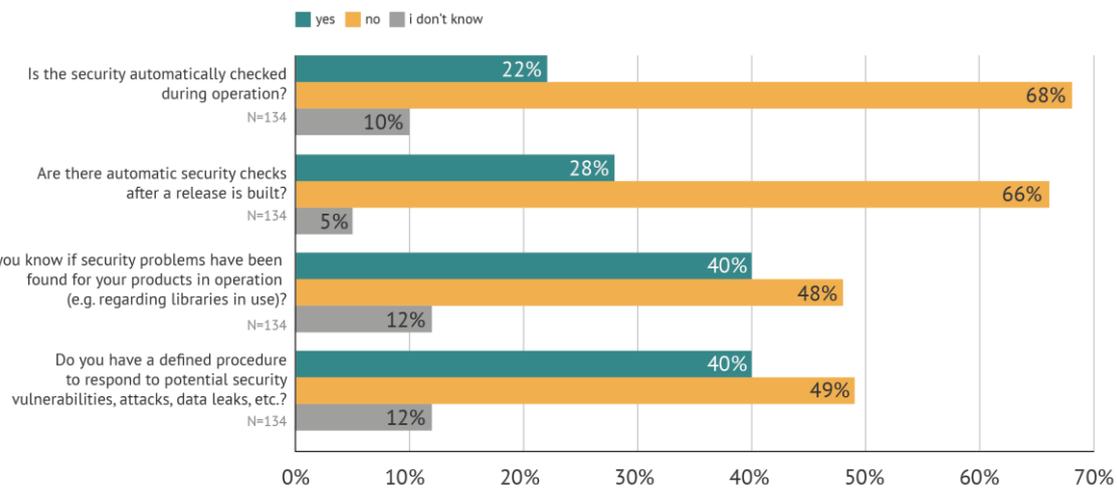

*Figure 15: Consideration of security in the discipline of operation*

To obtain further information on the procedure in a security incident, we asked the managers and product owners about this. One-third state that they have no processes and responsibilities or are not aware of them. At one quarter, the processes and company-wide responsibilities are primarily regulated. Neither the managers and product owners nor their developers have concrete responsibilities within a Product Security Incident Response Team (PSIRT)[4]. The processes and responsibilities are only partially regulated or known in the remaining managers and product owners, with reference usually only made to a data protection officer. However, this role is classically not responsible or accountable for security and, thus, not trained in this regard. The personal involvement of the manager and product owner in a security incident resolution process is rare overall.

"Mr. [...] can answer that for sure. That is our data protection officer, and he knows the process. I am not responsible there and honestly don't know the process specifically. "

Overall, the statements made by the managers and product owners are consistent with those of the developers that the processes and responsibilities for a security incident are not well defined enough in most cases. This is particularly important for reacting quickly to newly discovered and reported vulnerabilities. For example, dealing professionally with disclosed vulnerabilities can avoid bad press and restore lost trust.

### 4.3. Are the Current Processes Suitable For Developing Software Securely?

When asked about the overall process, 64% of the developers believe that their team does not invest enough time in secure software development (cf. Figure 16). Nevertheless, 62% of the developers believe that their overall development process and the associated tools - irrespective of security - are suitable for their needs. This apparent contradiction may be since secure software development has not gained high priority. As a result, the developers feel that they invest too little time in secure software development, but they do not see this as a problem because the current processes are perceived to work.

---

[4] A Product Security Incident Response Team is active within an organization around the occurrence of security incidents in products. This includes both prevention through appropriate awareness-creating measures and further education/training, but also in particular the rapid response and rectification of newly discovered vulnerabilities. To this end, the teams coordinate processes and communication channels within their own organization with the involvement of the reporting persons.

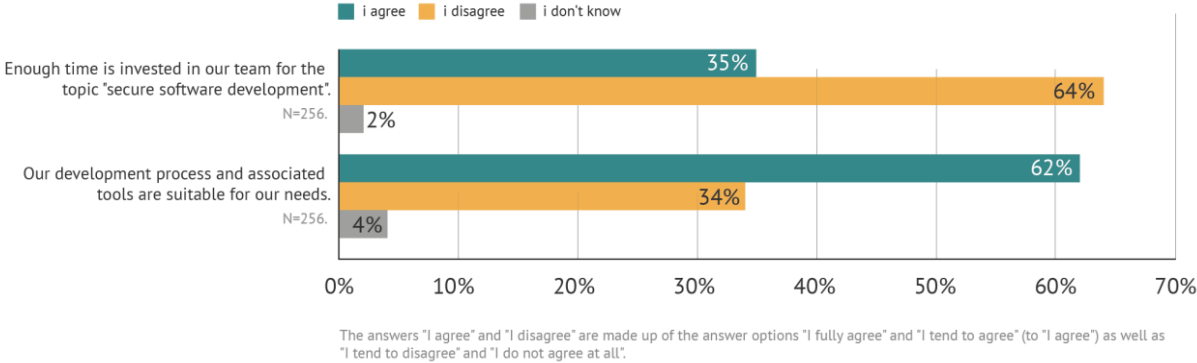

Figure 16: Suitability of current processes

A comparison of the various disciplines reveals a strong common trend (cf. Figure 17): Over two-thirds of the developers desire more precise and comprehensible processes (77% in the discipline of requirements engineering, 81% in design, 80% in implementation & testing, and 78% in operations). This suggests that the processes may work for a small proportion of the developers, but more precise and understandable processes are necessary across all disciplines.

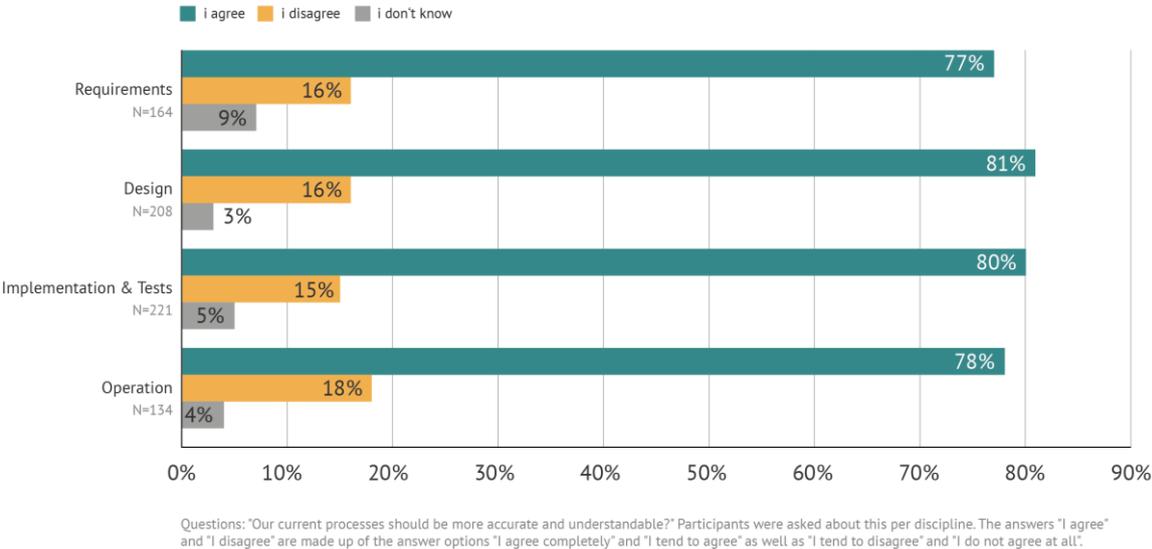

Figure 17: Assessment of the accuracy and comprehensibility of the processes in the respective disciplines

### 4.4. Interim Summary

The study participants agree that the current processes for secure software development and operation need improvement. First, the developers desire more understandable and more precise processes in all disciplines (~80% in each discipline). Second, the developers have the feeling that they invest too little time in secure software development (~60%). Finally, the developers use only very few measures (templates, standards, experts) for secure software development.

Although developers state that they pay attention to security, measures are rarely implemented to ensure systematic secure software development. Without these measures, however, such a guarantee is very unlikely. Thus, the developers have an inaccurate self-assessment when they say they pay attention to security. One possible explanation for this could be that most of those involved are not aware of the options available for secure software development.

It is also worrying that 20% of the developers admit to *not* paying attention to security during implementation and testing.

"Functionality before security. It's always more important that the button turns from green to blue instead of having another penetration test run over the application or the like. Because one brings money, makes the users happy, and the other, you don't get anything out of it at first."

# 5. Tools

> "So, I think it's perfect and important to have tools that we can use, both in the development environment but also much more so in our build pipeline."

In today's software development, tools are a decisive component for successfully implementing projects. They help developers to avoid errors or to find and correct them at an early stage. In addition, tools can automate manual and time-consuming activities, saving time and significantly reducing errors. Overall, the systematic use of tools along the development process leads to more efficient development while simultaneously increasing quality.

Some tools ensure the security of software products by supporting developers in systematically checking security requirements, generating secure code snippets, or detecting vulnerabilities. This support can significantly reduce the number of security errors and increase the application's security.

Besides all the advantages, however, the use of tools also has its limits: Various studies have shown that too many or unsystematically used tools lead to a long-term decline in software quality. This applies in particular to the aspect of security[5].

In the present study, we have investigated how tools are used in development today. In addition, we have determined what demand there is for such tools.

## 5.1. Do Developers Use Tools for Secure Software Development?

The developers state that they use tools along the entire development process, but to varying degrees of intensity (cf. Figure 18). Tools are used comparatively rarely in the disciplines before implementation (requirements engineering and design). In requirements engineering, 18% of the developers use tools to elicit and document security requirements. In design, 14% of the developers use tools to document security features. In both disciplines, ticket systems (e.g., Atlassian Jira) or Microsoft Excel are frequently used for documentation. In addition, the results of security analyses from previous releases are considered to ascertain further security requirements and take related properties into account in the design. Finally, 50% of the developers who use tools to collect or record security requirements also use tools to analyze their security requirements automatically. This enables them to find and correct errors at an early stage.

---

[5] https://www.heise.de/news/Studie-Mehr-Tools-weniger-Sicherheit-4799924.html

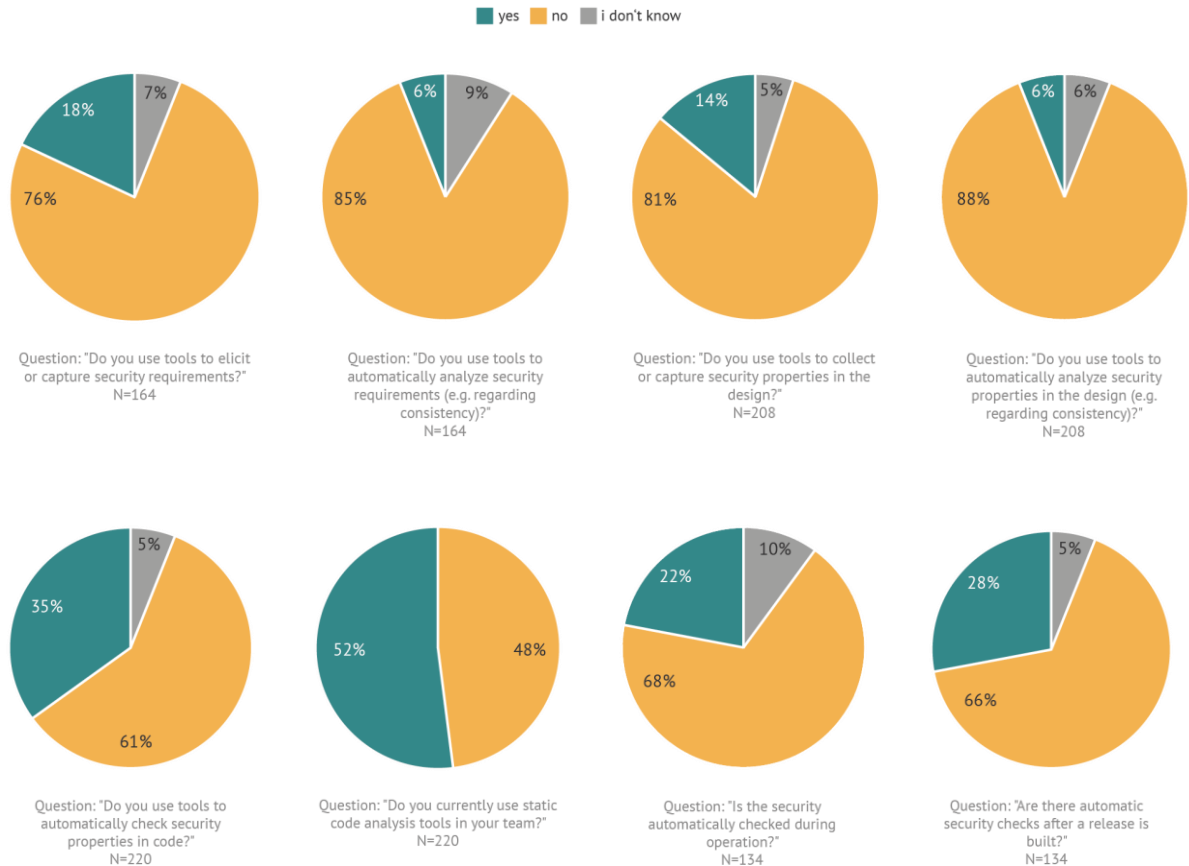

*Figure 18: Today's tool usage for secure software development*

During implementation, 35% of the developers use tools to check security properties automatically. In addition, a large proportion of the developers say they use tools for static code analysis (52%). Static code analysis tools can automatically analyze various properties in the code without executing it. For example, static code analysis tools can detect violations against the programming style or duplicate code. Developers most frequently use the static code analysis tools SonarQube and FindBugs. Both tools define security-relevant rules by default.

Most static code analysis tools can analyze simple security properties in the code. For example, if hard-coded credentials are in the code. But there are also special static analysis tools checking more complex properties in the code. These tools can, for example, check whether user input is verified before it is processed further or whether the developers use APIs correctly. However, only 50% of the developers who use static code analysis tools use these tools to analyze security properties.

A similar picture emerges for the use of tools during software operation. For example, 22% of the developers state that security is checked automatically during operation. In addition, 28% of the developers automatically analyze a release for possible vulnerabilities.

## 5.2. What Is the Opinion of Managers and Product Owners on Tools for Secure Software Development?

In general, the managers and product owners see a high demand for suitable tools for secure software development among their developers and are open to purchasing and introducing them.

The managers and product owners state that their developers use commercial and free tools. Furthermore, they agree that when it comes to tools, the focus is not on the costs but on the added value that can be achieved by using the tools. Furthermore, the managers and product owners have no negative opinions regarding free tools. In addition, the developers at most companies can procure and use their preferred free tools themselves.

When asked about their developers, most managers and product owners state that they only use tools they are convinced of. This can mean that different tools are used for the same purpose - especially if there are several free alternatives. This results in the risk that these tools are used in a less targeted manner. Only one product owner stated that free tools are regularly discussed, and guidelines for their use are created.

"The developers are free to use the tools they think are good. But, of course, there are rounds in which the developers present the used tools. Furthermore, they compare the advantages and disadvantages of the different tools."

The managers and product owners generally have a sufficient budget for acquiring charged tools. However, in contrast to free tools, they pay close attention to the added value that arises from using the tools. If the developers can sufficiently justify the added value, the interviewed managers and product owners support the acquisition. In addition, in the case of an investment, it is often demanded that the developers must also use these tools.

However, the managers and product owners reported that their developers rarely suggest procuring a paid tool. This contradiction may have several reasons:

- The selection of free tools is already suitable for the needs of the developers.
- Paid tools on the market are not suitable for developers. This could concern usability, performance, or missing functions, for example. Therefore, although there is a need and budget, no additional tools are purchased.
- The developers require improved tools, but they have not yet researched suitable tools.
- The developers, managers, and product owners do not talk about tools or talk about them only rarely. But, unfortunately, this does not lead to an improvement.
- There is a lack of overview of the spectrum that can be covered with tools or understanding security issues.

"Again, I could only speculate because this question (editor's note: the need for more security tools) has not yet been brought to me by the developers. But I would guess that they would be very grateful for any form of support. I assume that they also are interested in developing software that is as secure as possible. But if I'm honest, I'd have to say I haven't been asked such a question by the developers yet. "

### 5.3. What Are the Needs of the Developers?

When asked about the development process and the associated tools, almost two-thirds (62%) of the developers believe that these are suitable for their needs. However, one-third (34%) disagree with this statement (cf. Figure 19). In a second question, we asked the developers whether they have a suitable collection of tools for developing software securely. Here, only 40% state that this is the case, while most developers (56%) deny this. Thus, according to the developers, there is a clear need for more or better tools.

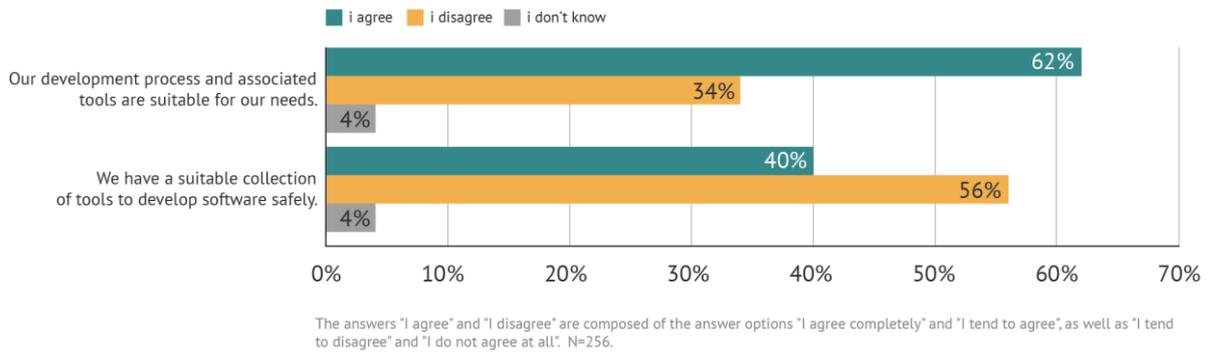

*Figure 19: Suitability of current tools for secure software development*

However, the lack of adequate tools differs in the individual development disciplines. While the need for more or better tools is high in the two disciplines' requirements engineering and design (56% and 64% respectively, cf. Figure 20), 72% believe that better tools for implementation would help complete the tasks better.

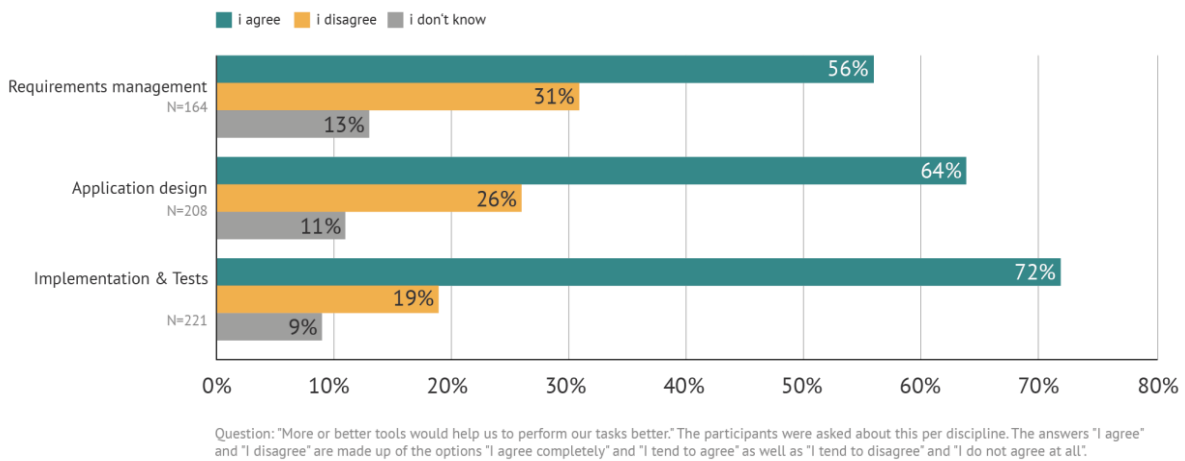

*Figure 20: Need for tools for secure software development*

## 5.4. Interim Summary

Tools are an essential component of successful, secure software development. The developers, managers, and product owners also see the importance of this topic. However, it has been shown that the distribution and use of tools are shallow, especially in the early development disciplines. Even during implementation, a large proportion does not use tools for secure software development. As a result, many errors are discovered late in development or after release, necessitating expensive and time-consuming repairs[6]. In addition, the risk of security incidents occurring in productive use increases.

Another result of the study is that the developers have a high unmet need for tools. However, according to the managers and product owners, there is a sufficient budget to purchase tools.

When using free tools, more attention should also be paid to ensure coordination occurs within the team. The tools can be used in a beneficial way and integrated well into existing processes.

---

[6] Quelle: NIST Planning Report 02-3, "The Economic Impacts of Inadequate Infrastructure for Software Testing", 2002, Page 5-4.

# 6. Security Expertise

> "You can never have enough competence. Learning never stops. "

The security competence of all participants (developers, managers, and product owners) is a crucial building block for systematically addressing security during development. We, therefore, asked all participants in the study for a self-assessment and asked them which competencies they have today, who should have which competencies, and whether the team's competencies are sufficient today.

### 6.1. How Do the Developers Rate Themselves?

To obtain a self-assessment of the developers' security competence, we divided the measures for secure software development into ten topics[7] along the development process (cf. Figure 9) and assigned these to the disciplines of requirements engineering, design, implementation & testing, and operation. The developers' task was to assess whether they were familiar with the topics and their concepts and, if so, whether they had gained practical experience in these topics. Thus, the survey aims to determine whether the developers have basic skills and experience and not whether they are experts in the respective subject topics.

Figure 21 shows the self-assessment results sorted by the proportion of practical experience (column 4). The best-known topic with practical experience is input validation, which should always be performed when reading in data (61%). Since this topic area also includes the significant attack type injection[8], this comparatively high value is encouraging. However, the developers have significantly less practical experience in all other areas. In patch management, for example, the figure is only 42%. In the remaining areas, between 36% and 21% of the developers have practical experience.

Concerning practical experience, it is striking that all four topics of the implementation discipline and the pen testing topic can be found in the upper half of Figure 21. In contrast, all other topics (everything that occurs before or after implementation) can be found in the lower half of the table. Thus, we conclude that security is granted more priority and time during the implementation than in the other disciplines. However, it would be necessary for the different disciplines to be given more priority and time for security for secure software development.

The pent test is a widely used measure to check the security of a software product before a release. Thus, the fact that the pen test is also in the upper half of the table is in line with our expectations. However, a pen test also has many limitations: It can only detect bugs and not show their absence like all testing procedures. In addition, its execution is usually limited to a few days (attackers do not have this limitation), and it does not investigate whether the requirements or the design already contain a vulnerability. Thus, all ten topics are relevant and not only the pen test.

---

[7] These topics are not fully comprehensive but cover a very large spectrum of secure software development.
[8] Among other things, Injection is ranked as the highest risk for web applications by the OWASP Foundation: https://owasp.org/www-project-top-ten/

| Subject area | Subject area not known | Subject area and concepts known, no practical experience | Subject area and concepts known and practical experience | Subject area at least known | Discipline in the process |
|---|---|---|---|---|---|
| Input check (sanitization) | 14% | 26% | 61% | 87% | Implementation |
| Patch Management | 25% | 33% | 42% | 75% | Implementation |
| Penetrations-Tests (Pen-Tests) | 18% | 46% | 36% | 82% | Test |
| Correct use of cryptography libraries | 22% | 43% | 36% | 79% | Implementation |
| Defensive Coding (Writing secure code) | 27% | 38% | 36% | 74% | Implementation |
| Design of secure architectures | 25% | 45% | 31% | 76% | Design |
| Security requirements and security test cases | 27% | 44% | 29% | 73% | Anforderungs-management |
| Incident Response (Behavior in the event of security incidents) | 29% | 45% | 26% | 71% | Operation |
| security-specific code reviews | 31% | 45% | 24% | 69% | Test |
| Carrying out a Threat analysis | 27% | 52% | 21% | 73% | Requirements management |

The topics are sorted in descending order according to the proportion of practical experience (column 4). The additional column "at least known" shows the percentage of developers who know the subject area and its concepts - regardless of practical experience (the respective value corresponds to the sum of the values from columns 3 and 4).

*Figure 21: How do developers rate their security competence per subject area?*

Column 5 of Figure 21 shows the extent to which each topic is at least known (this results from the sum of columns 3 and 4). This means that all topics are at least known by at least 69% of the developers. The input and pen tests achieved an outstanding 87% and 82%, respectively. Overall, it can be said that most of the topics have a good level of awareness and that no topic has a significant drop in awareness.

Figure 22 shows a second evaluation of the developers' self-assessment. Here, we have counted how many topics are unknown to him/her, known but without practical knowledge, or known and with practical knowledge. The result is that the developers have very different competencies. When looking at the extremes, 2% do not know any topics at all, and only 3% have practical experience in all topics. On the other hand, 39% of the developers know all topics - conversely, 61% do not know at least one topic.

We divide the number of developers roughly into three groups. Group 1 consists of about a quarter of the developers (23%). This group is not familiar with five or more topics and has only minimal practical experience in the few familiar topics. Group 2 comprises approximately 41%. They know five or more subject areas and have practical experience in some of them, but also do not know some subject areas. Finally, Group 3, which comprises about one-third of the developers, has practical experience in at least five of the subject areas and knows almost all of them.

In our opinion, all developers should know almost all topics. One reason for this is that all developers in the agile meetings (planning, review, retro) should understand what is being discussed and be able to have their say so that there can be an exchange of opinions and critical questions. Furthermore, we believe that all developers should have practical experience in many topics. One argument for this is that developers today are typically active in several or even all disciplines and, therefore, need practical experience to be able to consider the aspect of security. In particular, Group 1 requires an expansion of competence to be able to guarantee secure software development.

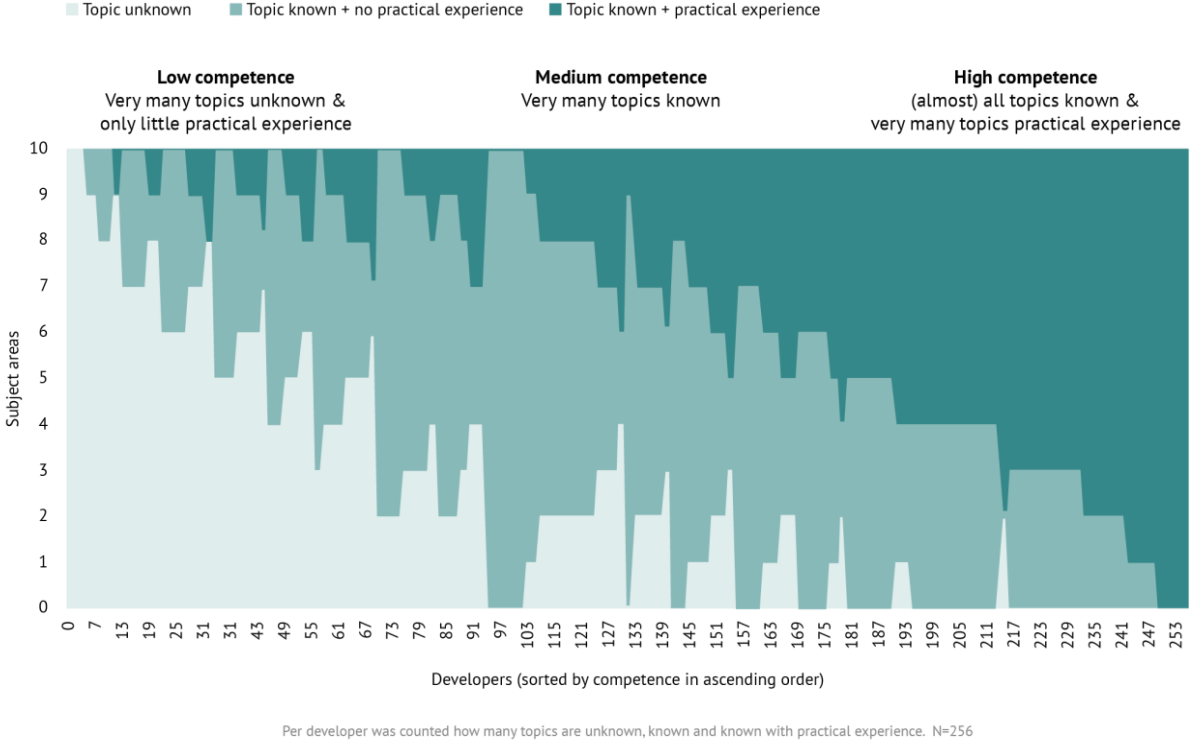

Figure 22: Awareness of the topics per developer

## 6.2. What Are the Competencies of the Managers and Product Owners?

In our opinion, all managers and product owners need essential security competencies. For example, both should know the laws and standards relevant to them, understand what security encompasses (including the distinction between data protection and security), and know their respective tasks and duties concerning security (including risk management). In addition, we believe that the manager must systematically support its employees in the further development of their security competencies. For the product owner, we see, among other things, the tasks of defining security requirements for the product and having it explained to them in the review meeting how security is guaranteed.

Our interviews showed that security is insufficiently known to the managers and product owners. Only for the discipline of implementation and testing does prior knowledge often seem to be available - especially for the topics of cryptography or encryption and pen tests. However, other methods and measures aimed more at the disciplines of requirements engineering, design, and release & operation are mentioned only very sporadically.

The managers and product owners frequently address data protection topics and the GDPR. However, some managers and product owners are unaware that security is about data protection and security. In addition, less than one-third of all the managers and product owners see the topic of security as part of their area of responsibility. Our conclusion, therefore, is that the current competencies of most managers and product owners are relatively low.

"I realize that IT security is a topic where I don't feel quite secure at all and, on the other hand, I am quite aware of the relevance of security and, yes, I realize that further competencies would do me good."

We then asked the managers and product owners whether they would like to have more security expertise. Two-thirds would like this - a few of them explicitly state that they do not feel competent enough. However, 20% of the managers and product owners say that they do not need any further expansion of competencies since, in their opinion, they have sufficient competencies to perform their tasks. On the other hand, a manager and a product owner state that they do not need any competence in secure software development, as this is not relevant for fulfilling their tasks.

Since - as mentioned above - all managers and product owners should have an essential security competence and our analysis has concluded that an expansion of competence appears necessary, it is gratifying that most managers and product owners also see it this way. This willingness should therefore have a positive effect on measures to expand competencies.

"I would have to get familiar enough to determine what the relevant issues are."

## 6.3. Should Every Single Person on the Team Have a High Level of Security Expertise?

It would be desirable if all the people in a team had a high level of competence in secure software development. But how do the developers judge this? Our survey results that 70% agree with this statement (cf. Figure 23). This is a surprisingly high value for us since it means, on the one hand, that most developers demand high competencies not only for a few but for everyone in the team. And on the other hand, the clear majority expects basic knowledge and high competencies from their colleagues. In summary, there is explicit agreement among the developers.

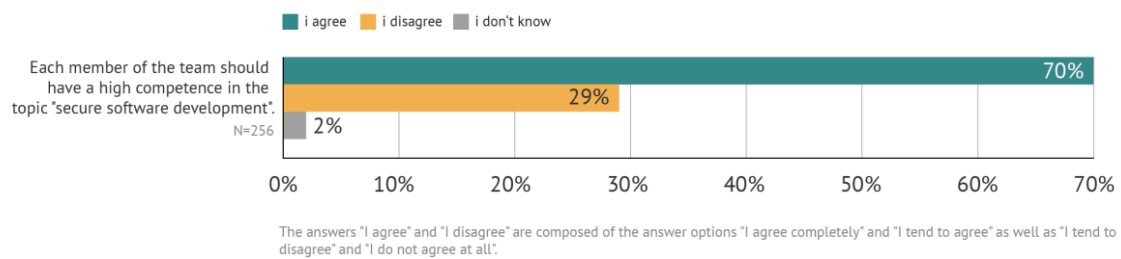

*Figure 23: Opinion of the developers regarding the competence of the team in the area of SSE*

## 6.4. Do the Developers Consider the Competencies of Their Team to Be Sufficient?

Almost two-thirds of the developers believe that the current competencies of their team are not sufficient for appropriate security requirements engineering, secure designs, and secure implementations, or secure operation (cf. Figure 24). Thus, the developers consider their competencies inadequate in all disciplines.

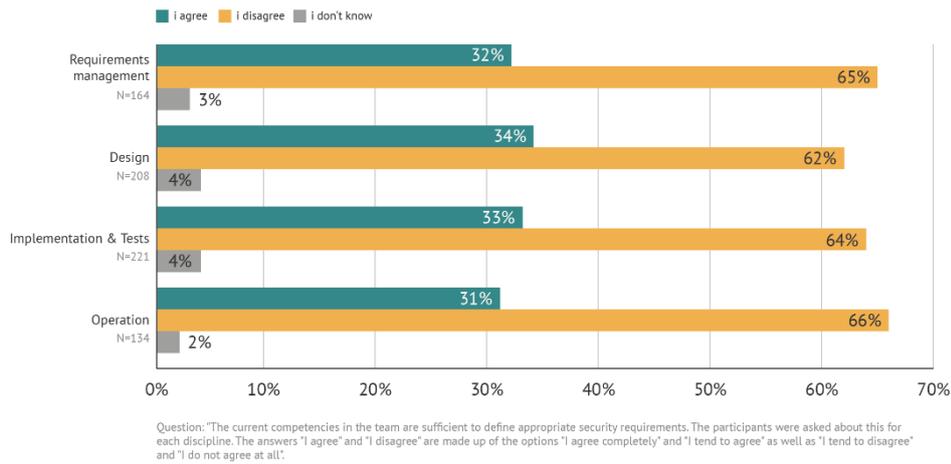

*Figure 24: Developers' assessment of current security competencies in the team*

In addition, we asked all developers whether they would like to receive further training from experts in secure software development. A clear majority of 89% confirmed this statement (cf. Figure 25), while the remaining 11% would not like any further training by experts[9]. Therefore, we conclude that almost all developers consider more competencies necessary for their teams and themselves.

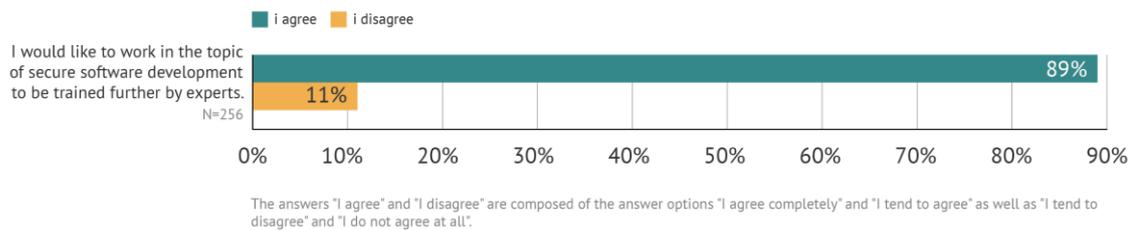

*Figure 25: Further training needs of the developers by experts*

## 6.5. What Are the Expectations of Managers and Product Owners for Their Developers?

We asked the managers and product owners which competencies they expect from their developers concerning secure software development. The result: Just over half of the managers and product owners named one to a maximum of four activities related to the discipline of implementation. In total, the managers and product owners mentioned the following six aspects concerning implementation:

- Encryption
- Secure communication
- Login/Passwords/Hashing
- Review of 3rd party libraries
- Input check
- Restriction of process authorizations

---

[9] 25% of the developers who do not want to be trained by experts in this topic state elsewhere in the questionnaire that they only prefer self-study.

Encryption and secure communication are mentioned most frequently (by around a third of all managers and product owners).

However, regardless of the discipline, 20% of the managers and product owners expect the developers to continue their education on their own (including self-study). One manager mentions a company guideline for secure software development and expects developers to have the necessary competencies; on the other hand, one-third of managers and product owners do not have a list of competency requirements.

> "I expect my software developers to have some basic understanding of security."

In addition, the managers and product owners stated three requirements, respectively:

- Have data protection competence
- Know the current state of the art
- Use provided tools

All other mentions are individual opinions; these are the correct behavior after a security incident (incident response), the ability to detect security incidents (incident detection), discovering security gaps using pen tests, coordinating security concepts with other teams, the ability to convince one's team and transfer responsibility to the team, having a security mindset, being able to decide for oneself what is essential for secure software, pointing out necessary security measures to the manager.

Finally, two product owners should be highlighted that differ significantly from the other managers and product owners: One product owner says that it is sufficient for developers to have data protection skills to handle customer data correctly. On the other hand, another product owner says that no requirements are deliberately defined for the competencies.

In summary, it can be said that the requirements of the managers and product owners for their developers are relatively small in scope, differ significantly, and there is significant overlap only concerning the implementation discipline. However, to fulfill their responsibilities and tasks concerning security, the manager and product owner should be able to define a comprehensive list of requirements that encompasses all disciplines. We see one primary reason why this is not currently the case in the inadequate security competence of the manager and product owner, which we discussed in detail in Section 6.2.

## 6.6. How Do the Managers and Product Owners Rate the Competence of the Developers?

More than half of the managers and product owners believe that their developers need to expand their skills in secure software development. Just under 20% of the respondents believe that at least a slight expansion of competencies or an expansion of competencies is necessary for a few developers. Two managers and product owners believe that no further expansion of competencies is essential since they assume that the developers already have sufficient competencies. Therefore, the conclusion is that, according to the managers and product owners, most developers need to expand their competencies in security.

> "I believe that we need to develop and build competence in secure software development."

We also asked the managers and product owners whether their developers would like to see their security expertise expanded. More than half of the managers and product owners estimate that only a few developers would like to see their expertise grew. The remaining managers and product owners are divided into three groups of roughly equal size: the first group believes that their developers do not want any further expansion of competencies. The second group states that their developers are actively seeking competence enhancement. On the other hand, the third group does not know whether its developers want to expand their competencies. Overall, the managers and product owners believe that some, but not all, developers would like to see an expansion of competencies. However, this contrasts with the findings in Chapter 6.4 where 89% of the developers state that they would like to receive further training from experts.

## 6.7. Interim Summary

This chapter has shown that the developers' competencies are often too low and very diverse. For example, the developers' self-assessment shows that they are unfamiliar with the topics and have very little practical knowledge. Two-thirds of the developers also think that the current skills of their team are not sufficient to develop or operate software securely. However, it is encouraging that more than two-thirds of the developers believe that all team members should have a high level of competence.

Many managers and product owners do not have sufficient competencies to fulfill their tasks. Most managers and product owners also see it this way, as two-thirds would like to have more competencies themselves.

The manager and product owner have few requirements for their developers. Furthermore, it has been shown that there is no uniform or majority opinion regarding the required competencies. Most of the managers and product owners refer to the discipline of implementation. The other disciplines seem to be less in focus for the managers and product owners. However, more than half of the managers and product owners think that their developers need to expand their competencies in secure software development.

## 7. Competence Expansion

"I believe that this topic is so extensive and, above all, so dynamic that it is not enough to give developers somehow training and then expect them to be able to create security-relevant software. [...] This is an ongoing process, and it has to be looked after permanently."

This chapter looks at the current development of companies' competencies. First, we discuss the training on offer in German-speaking countries, the extent to which developers inform themselves about new potential security vulnerabilities, and whether they attend meetups and conferences. Second, we explain how managers and product owners support developing their developers' skills and what future training formats developers would like to see.

### 7.1. Is the Training on Offer in German-Speaking Countries Satisfactory?

Less than half of the developers (40%) are aware of the training courses available in German-speaking countries on secure software development (cf. Figure 26). One possible reason for this is that the developers have not yet actively searched for (German-language) training courses. At the same time, there is a suspicion that the existing security training courses in the German-speaking world have been advertised too little or too infrequently or have not been noticed by the developers.

Of the 40% of the developers who are aware of the offering, almost two-thirds state that they are not satisfied. This is, therefore, a clear indication that the existing training courses known to the respondents do not fully cover the needs of the developers. Whether this is due to the content or the training format must be discussed in future studies.

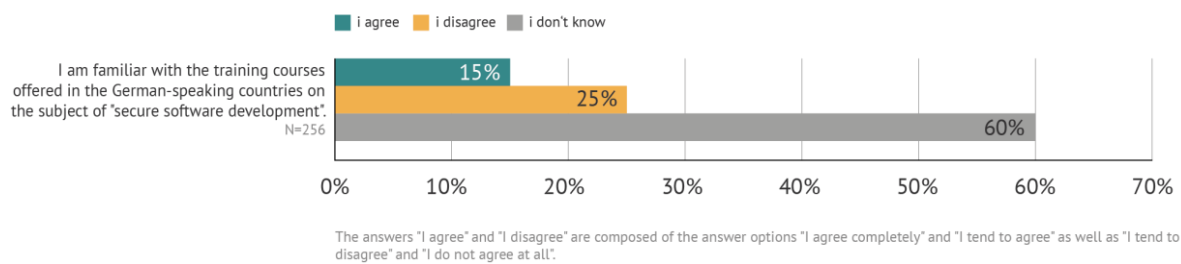

*Figure 26: Satisfaction of the developers with the training on offer in German-speaking countries*

Most managers and product owners are not familiar with the training courses offered in the German-speaking world concerning secure software development and are therefore unable to evaluate them. Occasionally, the managers mention internal company training opportunities, but according to the managers and product owners, these typically have little or no focus on security. One manager also raises the issue that too little competence in secure software development is taught in training and university teaching.

Furthermore, the managers and product owners often note that security topics are not preferred by developers when selecting the appropriate training courses. Instead, a new framework, a new tool, or a different programming language seems to be more popular than security. As described in Section 6.2, we believe that it is the responsibility of the managers, in particular, to do more to promote the development of security skills among their developers.

"The developers' primary desire for knowledge is certification and training in new technologies. Security is not the top priority."

## 7.2. Do Developers Keep Themselves Informed About New Potential Security Vulnerabilities?

Concerning today's competence development, we asked the developers whether they regularly inform themselves about new potential security vulnerabilities (for example, in the libraries and frameworks used) and where they obtain the relevant information. 55% stated that they regularly inform themselves (cf. Figure 27). In our opinion, almost all developers should periodically inform themselves about new potential security vulnerabilities - so we classify this value as relatively low. Of course, it would be legitimate for developers to use tools to automatically check whether there are potential security vulnerabilities in the product they are currently developing or operating.

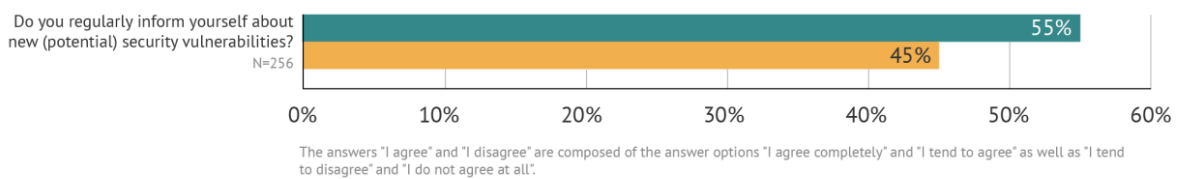

*Figure 27: Proportion of the developers who regularly inform themselves about new potential security vulnerabilities*

We then allowed the developers to name their sources of information. IT news websites are the primary source of information for the respondents (Heise and Golem were cited most frequently). Two-thirds of respondents (64%) cited sources from this category. Newsletters or mailing lists were named second most often (13%), and social media (Twitter, Github: 11%) third most frequently. Other mentions in the high single-digit percentage range were the BSI, blogs, CVE, and manufacturer websites. All in all, it can be said that the editors of IT news are currently held in high trust and are expected to report on critical gaps on time.

## 7.3. Do Developers Attend Meetups and Conferences on Security Topics?

Meetups and conferences on security topics are another way of expanding expertise. However, 77% of the developers stated that they do not regularly attend such events (cf. Figure 28). This figure is in line with the results of our survey of managers and product owners, as most of them state that their developers never or only very rarely attend meetups and conferences on security topics. Typically, the managers and product owners see it as the developers' duty to find suitable events for themselves. Thus, as soon as a proposal is submitted (within Europe), it is usually accepted.

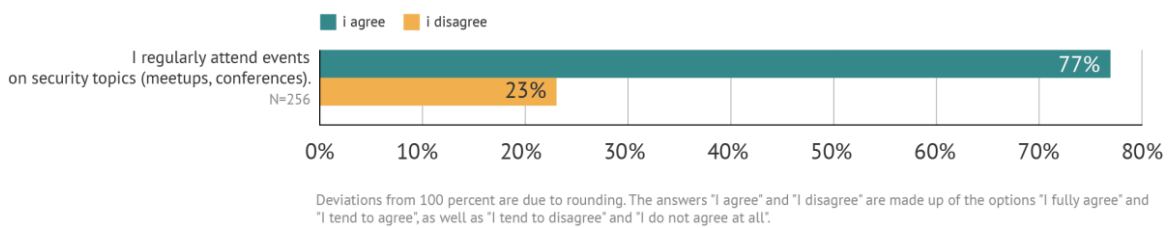

*Figure 28: Regular attendance of events on security topics by developers*

## 7.4. How Do Managers and Product Owners Support the Development of the Developers' Skills?

To increase the competence of their developers concerning secure software development, most managers and product owners enable them to attend internal and external training courses and conferences. Other measures are also mentioned in isolated cases: For example, a developer breakfast was mentioned, where developers from one area meet for breakfast and discuss a wide range of IT topics, with the topic of security frequently coming up. Another measure mentioned is that one person per team is set up as a so-called security champion, who takes the topic of security into account comprehensively in software development and involves colleagues and the product owner in this (e.g., for the definition and acceptance of security requirements).

"There are, of course, opportunities for all developers to continue their education during their working hours at conferences or within the scope of projects here at the company. A lot has already happened in this area, and security is always one topic."

Nearly all managers and product owners also stated that their developers are given time to study on their own about security, among other things, and that they welcome it if their developers take up this offer. However, many managers and product owners do not know whether this opportunity is used for security topics. In addition, the managers and product owners do not recommend these topics. Only one product owner states that he does not think it makes sense for his developers to continue their education in self-study since a trainer is urgently needed for this.

"You will always get the time from me to study on your own."

Except for one product owner, all interviewed managers and product owners stated that they provide sufficient financial resources to enable their developers to take part in internal and external training measures. Typically, the budget is used for internal and external training and allows developers to attend conferences. In most cases, all requests are granted by the managers and product owners as long as they consider the event meaningful in terms of content.

There is no fixed budget for further training measures in most cases, but the money is granted on request. However, one manager states that all developers have a limited budget of 1,000 euros per year. If this limit is not exceeded, the developers may decide completely independently on which further training they invest the money. According to the manager, this model is very well accepted by the developers, e.g., for continuing education in online courses. Another manager states that it has deliberately not defined a budget limit. This is because it does not want to restrict its developers concerning further training - the topic of security would always be approved.

"We have the unique feature that we do not have any budget limits for such training courses. Training courses that make sense can be attended. The term security would always be classified as meaningful by us. "

We also asked the managers whether they had formally defined security competencies in their job profiles. Almost all managers stated that this is not the case. One manager does not know whether this is the case. Another manager says that it is, but that security competence is only decisive for applications for extremely security-relevant roles. However, one manager states that it is not defined in the job profile adds that this topic is nevertheless addressed in the interview.

In summary, it can be said that the managers and product owners typically enable their developers to participate in internal and external training courses and conferences - the budget is not a limiting factor here. In addition, many managers and product owners enable and desire self-study for competence development but do not know whether this offer is used. The managers and product owners leave it up to their developers to decide whether they want to expand their security expertise. Only the definition of a security champion is a targeted measure on the manager and product owner to develop security competence. Since it became clear in Section 6.6that almost all managers and product owners favor a significant expansion of security competence among their developers, they should drive this forward more systematically and demand it more strongly.

## 7.5. Which Training Format Would the Developers Like to See?

As already reported in Chapter 6.4, 89% of the developers state that they would like to receive further training from experts in the area of secure software development (cf. Figure 25).

We also asked the developers which options they preferred for further training, with multiple answers possible (cf. Figure 29). The self-study was named most frequently (62%). Just over half of the respondents (52%) said that a workshop lasting a maximum of one day or a seminar lasting 2-3 days would be useful. Approximately one-third (32%) find an online training course with a trainer useful, and one in five find a seminar lasting 4-5 days useful.

In conclusion, it would make sense to offer different formats for teaching the same or similar content to address the diverse learning preferences.

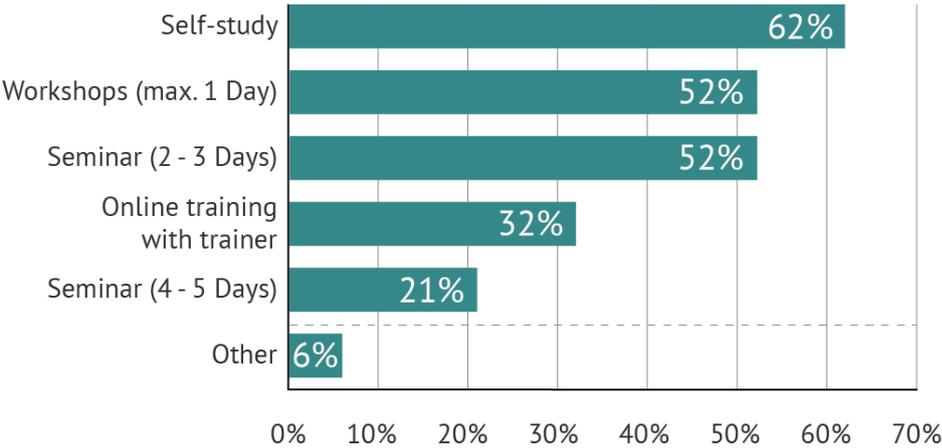

Question: "Which opportunities for continuing education do you prefer?"
Multiple answers possible. N=256.

*Figure 29: Which training formats would developers like to see?*

## 7.6. Interim Summary

Overall, it can be said that the current expansion of competencies needs to be improved in several areas. For example, most developers, managers, and product owners are not aware of training offered in German-speaking countries. However, most of those involved stated in Chapter 6a need to expand competencies in security. In addition, of the developers who are aware of the training on offer, two-thirds are not satisfied with it - here, too, there is a need for action.

Only slightly more than half of the developers regularly inform themselves about potential security vulnerabilities, and 77% of the developers say they do not attend meetups and conferences on security topics. Therefore, we recommend an increase in activities.

Another finding is that the managers and product owners enable their developers to continue their education through training courses, conferences, or self-study. However, it became clear that the managers and product owners do not systematically promote or demand the development of security competence among their developers. The exception to this is the measure to appoint a security champion, which was reported by only a few managers and product owners.

Ultimately, it has become clear that different formats should be offered for teaching competencies to address developers' different learning preferences.

## 8. Awareness of All Involved

"Security is a huge issue from my perspective, and you think about it in between. But those are thoughts I push away."

The previous chapters have shown that the quality of the development processes for secure software development, the dissemination of suitable tools, and security competence are usually low or mediocre. A possible cause for this could be insufficient awareness of individual or even all participants. An inadequate awareness would have the consequence that the risks of unsafe products are not conscious. Due to it, no efforts are made to improve the processes, use more or more suitable tools, or develop Security authority. In the following, we deal with how all participants are aware concerning security. We first examine the awareness of managers and product owners and then the developers' awareness.

To answer the question, we divide awareness into three categories:

- **No awareness:** A person is not aware of secure software development if he or she is unfamiliar with the topic or does not understand its benefits. Naturally, people in this group see no need to expand competencies, improve processes, or use tools.

- **Low awareness:** A person who is aware of secure software development and understands its benefits has low awareness. Essential competencies are sufficient for a topic to be known. These competencies can already be acquired in rudimentary form through one of the GDPR training courses in companies today. Only data protection topics are dealt with in such training courses, but security is not addressed. However, knowledge of both topics is necessary for comprehensive awareness.

- **Comprehensive awareness:** An awareness of data protection, data security, and security is essential for a broad awareness of the topic of secure software development. Therefore, people in this group are aware of the risks that a case of data loss, theft, or manipulation can pose and see a clear need for action if insufficient competencies are available, processes are incomplete, or tools are missing or incorrectly used. In addition, comprehensively aware persons can clearly distinguish between data protection and security and know how to apply them. Furthermore, we define for persons in this category that they can name their tasks and responsibilities in the context of security.

### 8.1. Are the Managers and Product Owners Aware?

Chapters 4 and 5 showed that the managers and product owners approach security in relation to the development processes and tools rather unsystematically. However, a need for action is necessary in each case. This indicates a relatively low level of awareness. The only exception is that the managers and product owners are always willing to invest money in paid tools for secure software development. Naturally, the developers desire this, and there is a meaningful benefit.

As already reported in Section 6.2, we believe that most of the managers and product owners have relatively low-security competencies. In addition, they often state that security is important but fail to mention any measures to ensure it. Overall, this leads us to conclude that only a few managers and product owners are comprehensively aware.

In addition, we showed in Chapter 7 that the managers and product owners are currently approaching the development of their developers' competencies rather unsystematically. However, there is a clear need for more security competencies on the part of the developers. This is, therefore, a further indication that the UK and product owners have only a low level of awareness at most.

"I think if we had a security issue, we would tell individuals, 'Get fit for this.'"

Almost all managers and product owners whose products are not publicly accessible but are used internally within the company or internally at their customers are at best only slightly aware since they typically regard the topic of security as almost insignificant. Often the firewall is mentioned here, which is supposed to provide sufficient protection. However, many respondents are not aware of the danger of an internal perpetrator or that a firewall does not provide adequate protection.

"A team that works purely on indoor applications indeed handles application security a little differently than a team that builds software that also runs on websites, so is accessible even from the Public Internet, and is also correspondingly exposed to attacks daily."

We also examined the extent to which customer behavior contributes to raising the awareness of the managers and product owners. Most of the managers and product owners stated that many customers generally request security, and in some cases, even by all customers. According to the managers and product owners, those customers who request security analyses or carry out an audit themselves often come from large companies in which fixed security guidelines are defined. But even these guidelines are not perfect: For example, we were told that a pen test is required for the first release, but not for all subsequent releases, even if the product is developed further over several years. In the case of customers from small companies, these security guidelines are relatively rare, so the security requirements depend, according to the interviewed managers and product owners, firmly on the security understanding of the respective project participants of the customer. Overall, some customers do not demand security at all or only want to invest time in security when the project is awarded and in the further course of the project. However, the managers and product owners almost always view this customer behavior critically. This is, therefore, an indication of a low to a comprehensive level of awareness on the part of the managers and product owners.

"The awareness of customers is improving, no question about that. But since our requirements are also continuously increasing, I would say the understanding is only getting better in parts."

On the other hand, most managers and product owners state that they raise the security awareness of their customers or explicitly offer to take security into account in the project - but they do not manage to convince their customers of this regularly. In our opinion, this is a sign that the sales department in particular and the managers and product owner need further training to convince their customers of the importance of security. Nevertheless, this speaks for a low to comprehensive awareness of the managers and product owners, as they are aware of this problem.

"So if you ask our customers now 'Should the applications be secure?', of course, no one says 'No, they shouldn't,' but of course they should be secure. But concrete measures are shockingly rare to sell to our customers."

Fewer than one-third of the managers and product owners believe that they do not even offer security because their customers have no need for it or do not understand its effort. Therefore, these managers and product owners are at best only of a low level of awareness.

## 8.2. Are the Developers Aware?

A clear indication of at least a low to a comprehensive level of awareness among a majority of the developers is that during requirements engineering, design, and implementation, at least two-thirds of the developers state that attention is paid to the issue of security (cf. Figure 30). Concerning the release, we did not ask whether attention is paid to security but whether there is a final security review before the release and automatic security checks after a release. However, the percentage of developers who answered at least one of the two questions positively was only about one-third. Therefore, it is still reasonable to suspect that most developers are not comprehensively aware concerning the release of the software.

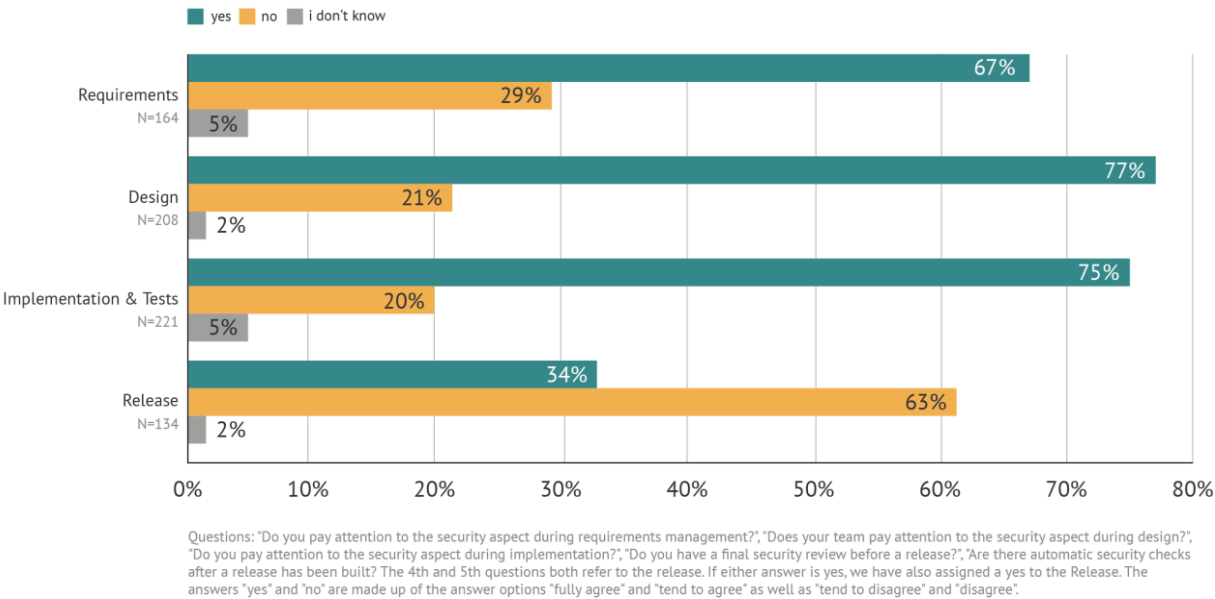

*Figure 30: Is attention paid to security in the different phases of software development?*

However, as explained in Chapter 4, there is a contradiction between the developers' assessment that attention is paid to security in the three disciplines of requirements engineering, design, and implementation & testing, and the statement that specific templates, standards, processes, and tools rarely exist and that there is rarely a security expert. This is a clear indication that most developers are not comprehensively aware. In addition, 20% of the developers state that they never analyze their software concerning security (cf. Figure 14 in Chapter 4). If they are not explicitly prevented from doing so by the product owner or customer, this group is also not comprehensively aware.

A decisive factor in raising awareness of security is the extent to which the developers have security expertise - in other words, whether they are familiar with the various topics and what practical experience they have in each of them. The higher the level of competence, the higher we rate the level of security awareness. As reported in Section 6.1 security competence of the developers is distributed very differently. The three categories of competence, low, medium, and high, are distributed roughly from 23 to 41 to 36. Developers with a medium or high level of competence are, in our estimation, slightly too comprehensively aware. We classify developers with low-security competence as having a low level of awareness. The four developers who did not know any of the ten topics are, in our opinion, not aware at all.

As described in Section 6.4, the vast majority of the developers (89%) in our survey expressed the desire to receive further training in the area of secure software development from experts. Thus, we conclude that they are aware of the topic's relevance and would like to develop more competencies. This is a strong indication that most developers are at least slightly aware. Regarding the remaining 11%, there are several possible reasons for not wanting this, e.g., disinterest in the topic, preference for self-study, or the assessment that their competence is high enough.

In addition, we asked the managers and product owners whether their developers are aware. The result is that one-third of the managers and product owners estimate that all developers within their teams are comprehensively aware. Another third of the managers and product owners, on the other hand, think that most developers are not at all or only on a low level of awareness. However, those individual members of the team are comprehensively aware to security. Typically, these individuals either have the explicit task of looking after security (e.g., the team's security champion) or having a personal interest in this. However, one-third of all interviewed managers and product owners suspect that even a low level of awareness is not present. Overall, we can conclude that at least half of their developers are not at all or only on a low awareness level according to the assessment of the managers and product owners.

"Of course, many developers know of security. They are also somewhat aware, as you usually hear about it in the media. Also, through conferences, where this is addressed. But I don't feel that this is practiced in day-to-day work. "

"About a third of the developers have a bellyache when they do things, and security is too unimportant. That doesn't divide over senior or junior either. That's a more personal inclination. "

## 8.3. Interim Summary

The managers and product owners have typically only a low awareness level. The predominant attitude is that security is necessary but not important enough to be addressed systematically and with high priority in processes, tools, and competence development. However, this would be necessary, as shown in the previous chapters. All managers and product owners are aware of data protection, primarily due to the GDPR, but rarely aware of the security or the risk of internal perpetrators when their products are used non-publicly. Security-aware and committed managers and product owners are also frequently confronted with a lack of understanding, resistance from superiors, or rigid processes.

Most developers are also only on a low awareness level, which is confirmed by the assessment of the manager and product owner. In addition, however, there is a minority of the developers who are either not at all or comprehensively aware. It is worth noting the contradiction in the inaccurate self-assessment of many developers, who believe that they are aware of the issue of security during the development and operation of their products, but often have no methods, tools, or experts and are, therefore, unable to systematically ensure security.

## 9. Conclusion and Effects

Using Germany as an example, our study has shown that ensuring security is a multi-layered challenge for companies. In the following, we briefly summarize the key findings of our study:

- Most managers, product owners, and developers have only a low level of awareness: they feel security is essential but do not act accordingly. Many managers and product owners lack awareness regarding security and internal perpetrators.
- Most developers have an inaccurate self-assessment on security: They think that they pay attention to the subject of security - but state that they do not have the appropriate measures, processes, internal experts, and tools.
- All participants in software development need more competence in the subject of secure software development. Although almost all participants in the study would like to see an increase in competence, this has so far only happened sporadically and unsystematically.

The current situation in software development harms the security of software products in the medium and long term. The combination of low awareness and inaccurate self-assessment on the part of all those involved in software development means that the current state is perceived as sufficient. No improvement (e.g., systematization of processes and expansion of competencies) is sought. However, the steady increase in attacks and the numerous security incidents in recent years show that ensuring security is becoming increasingly essential to avert[10] dangers for companies and customers. Thus, it is necessary to develop itself as an enterprise in secure software development continuously. In the following chapter, we provide recommendations for managers, product owners, and developers to improve the current situation.

---

[10] https://www.bsi.bund.de/DE/Publikationen/Lageberichte/lageberichte_node.html

# 10. Recommendations for Improving and Avoiding the Deficits

To improve secure software development, we recommend expanding the basic security knowledge of all developers, managers, and product owners. While developers should have extensive knowledge and practical experience, it is sufficient for managers and product owners to have a basic understanding of performing their tasks and duties concerning security.

In addition, we recommend that all involved parties raise each other's awareness so that there is a fundamental appreciation of security throughout the company. For example, the topic of secure software development should always be addressed openly - whether in the sprint review or concerning possible new tools or training courses. In this way, everyone can contribute to more secure software development.

Concerning processes and tools, no universally applicable solution makes sense for all companies, people, and products. We, therefore, recommend that developers, managers, and product owners jointly define how they want to ensure security throughout the entire development process.

We have summarized specific recommendations for individual roles below. These are intended as suggestions to counteract those revealed by the study.

## 10.1. Recommendations for Managers

- Together with the management and the product owners, define who is responsible for security issues. It is essential to determine who will ensure the long-term security of your products and what the processes and commitments are in the event of a security incident (keyword: PSIRT).
- You should know at least the ten security topics. This basic knowledge is relevant to accompany the security-specific further development of your developers and fulfill your responsibility concerning security. However, in-depth technical details are not necessary.
- When acquiring projects, make sure that there is room for security. Without appropriate budgeting, you will "pay for it" later.
- Raise awareness among your product owners and teams about security so that everyone understands the need to act together. Take advantage of existing training offerings or seek advice to clarify your individual needs.
- Define what basic security knowledge all product owners and developers should have - regardless of the specific project or product. Demand this in the medium term.
- Coordinate with your product owners and the developers whether advanced security competencies are necessary to develop the product. If necessary, derive further training requirements and corresponding measures.
- Motivate your employees to get further training in security topics.
- Consider whether having security experts (often called security champions) tasked with bringing security knowledge to the teams can be a valuable strategy for your organization. In general, we recommend that every team should have at least one security champion.
- Include money in the budget for training and tools related to security.

## 10.2. Recommendations for Product Owners

- If you are working with developers, you should know the ten security topics (cf. Chapter 6.1) to have a say (however, in-depth technical details are unnecessary).

- If you are a proxy product owner: Educate your customers about security risks and measures. Make sure your customers value and budget for the subject area.
- Raise awareness of security among your managers and team so that everyone understands the need for action.
- Define with your team how important security is for your product. Together, derive security requirements that your product must meet. Then, use security maturity models to develop your team and your product accordingly.
- Define requirement profiles for the developers of your product. Draw the attention of the manager to this.
- Demand security from the team during planning, review, and retrospective. Have the team explain how you will ensure security. Ensure you understand the explanation to judge for yourself whether the security is assured.
- Schedule and allow your team to address security issues in their sprints continuously.

## 10.3. Recommendations for Developers

- Know the ten security topics and understand how the corresponding measures can be implemented technically. Remember the measures during programming and gain practical experience in as many topics as possible: request time and budget for further training (internal and external) from your product owner or manager.
- Define which roles and tasks you and all other software developers have concerning security.
- Raise the security awareness of your manager and product owner so that everyone collectively understands the need to act.
- Get an overview of existing tools. Use tools to improve security in your development environment (IDE) and build pipeline. In addition, check whether tools for requirements engineering and design help you.
- Disseminate knowledge within the team, for example, on security basics, processes, and tools so that they do not have to act as a lone wolf.
- Think about how a hacker could attack or exploit your system - e.g., using Evil User Stories. Discuss this with colleagues and your product owner. Develop effective countermeasures.
- If your team does not yet have a person with high-security expertise (often called a security champion), you should discuss whether it makes sense to introduce such a role with the product owner.
- Ensure that for each tool you use, at least one person in your team has expertise in that tool and is the contact person for the team. Ensure that everyone else on the team knows the tool well enough to use it appropriately. Use internal or external training if necessary.
- Review your development process, especially concerning security. Then, improve and expand your process on an ongoing basis.
- Sort yourself into a security maturity model for agile teams to understand how good your team already is in security. Then, if necessary, derive measures to reach higher levels in the maturity model.

# Acknowledgments

This work is part of the research project "AppSecure.nrw - Security-by-Design of Java-based Applications" funded by the European Regional Development Fund (ERDF-0801379).

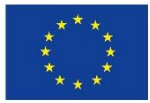 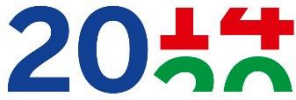